\documentclass[twocolumn,preprintnumbers,aip,amsmath,amssymb]{revtex4}

\usepackage{graphicx}
\usepackage{dcolumn}
\usepackage{bm}

\begin{document}
\title{Nanoscale magnetic field mapping with a single spin scanning probe magnetometer}
\author{L.~Rondin$^{1}$}
\author{J.-P.~Tetienne$^{1}$}
\author{P.~Spinicelli$^{1}$}
\author{C.~Dal Savio$^{2}$}
\author{K.~Karrai$^{2}$}
\author{G.~Dantelle$^{3}$}
\author{A.~Thiaville$^{4}$}
\author{S.~Rohart$^{4}$}
\author{J.-F.~Roch$^{1}$}
\author{V.~Jacques$^{1}$}
\email{vjacques@lpqm.ens-cachan.fr}
\affiliation{$^{1}$Laboratoire de Photonique Quantique et Mol\'eculaire, Ecole Normale Sup\'erieure de Cachan and CNRS UMR 8537, 94235 Cachan Cedex, France}
\affiliation{$^{2}$Attocube systems AG, Koeniginstrasse 11A RGB, Munich 80539, Germany}
\affiliation{$^{3}$Laboratoire de Physique de la Mati\`ere Condens\'ee, Ecole Polytechnique and CNRS UMR 7643, 91128 Palaiseau, France}
\affiliation{$^{4}$Laboratoire de Physique des Solides, Universit\'e Paris-Sud and CNRS UMR 8502, 91405 Orsay, France}

\begin{abstract}
We demonstrate quantitative magnetic field mapping with nanoscale resolution, by applying a lock-in technique on the electron spin resonance frequency of a single nitrogen-vacancy defect placed at the apex of an atomic force microscope tip. In addition, we report an all-optical magnetic imaging technique which is sensitive to large off-axis magnetic fields, thus extending the operation range of diamond-based magnetometry. Both techniques are illustrated by using a magnetic hard disk as a test sample. Owing to the non-perturbing and quantitative nature of the magnetic probe, this work should open up numerous perspectives in nanomagnetism and spintronics.
\end{abstract}

\maketitle

\indent The ability to map magnetic field distributions with high sensitivity and nanoscale resolution is of crucial importance for fundamental studies ranging from material science to biology, as well as for the development of new applications in spintronics and quantum technology~\cite{Kirtley2010,Romalis_NatPhys2007,Bogani2008}. In that context, an ideal scanning probe magnetometer should provide quantitative magnetic field mapping at the nanoscale under ambient conditions. In addition, the magnetic sensor should not introduce a significant magnetic perturbation of the probed sample.\\ 
\indent Over the last decades, different roads have been taken towards ultra-sensitive detection of magnetic fields including superconducting quantum interference devices (SQUIDs)~\cite{Kirtley2010}, semiconductor-based Hall probes~\cite{Kirtley2010} and optical magnetometers~\cite{Romalis_NatPhys2007}. Even though extremely high sensitivity has been achieved with these devices, their spatial resolution remains limited at the micron-scale. Prominent approaches to reach nanoscale resolution are scanning-tunneling microscopy~\cite{Heinze_Science2000}, mechanical detection of magnetic resonance~\cite{Rugar_Nature2004}, nanoSQUIDs~\cite{Yacoby_NanoLett2010}, X-ray microscopy~\cite{Chao2009} and magnetic force microscopy (MFM)~\cite{MFM}. Since the latter technique operates under ambient conditions without any specific sample preparation, it is now routinely used for mapping magnetic field gradients around magnetic nanostructures. However, besides introducing an inevitable perturbation of the studied magnetic sample owing to the intrinsic magnetic nature of the probe~\cite{Garcia2001}, MFM does not provide quantitative information about the magnetic field distribution. \\
\indent Here we follow a recently proposed approach to magnetic sensing based on optically detected electron spin resonance (ESR)~\cite{Chernobrod2005}. It was shown that this method applied to a single nitrogen-vacancy (NV)  defect in diamond could provide an unprecedented combination of spatial resolution and magnetic sensitivity under ambient conditions~\cite{Degen_2008,Taylor2008,Balasubramanian2008,Maze2008}. The principle of the measurement is similar to the one used in optical magnetometers based on the precession of spin-polarized atomic gases~\cite{Romalis_NatPhys2007}. The applied magnetic field is evaluated by measuring the Zeeman shifts of the NV defect spin sublevels. In this article we demonstrate {\it quantitative} magnetic field mapping with nanoscale resolution, by applying a lock-in technique on the ESR frequency of a single NV defect placed at the apex of an atomic force microscope (AFM) tip. In addition, we report an all-optical magnetic imaging technique which is sensitive to large off-axis magnetic fields, and relies on magnetic-field-dependent photoluminescence (PL) of the NV defect sensor, induced by spin level mixing.\\
\begin{figure}[t]
\includegraphics[width = 8.4cm]{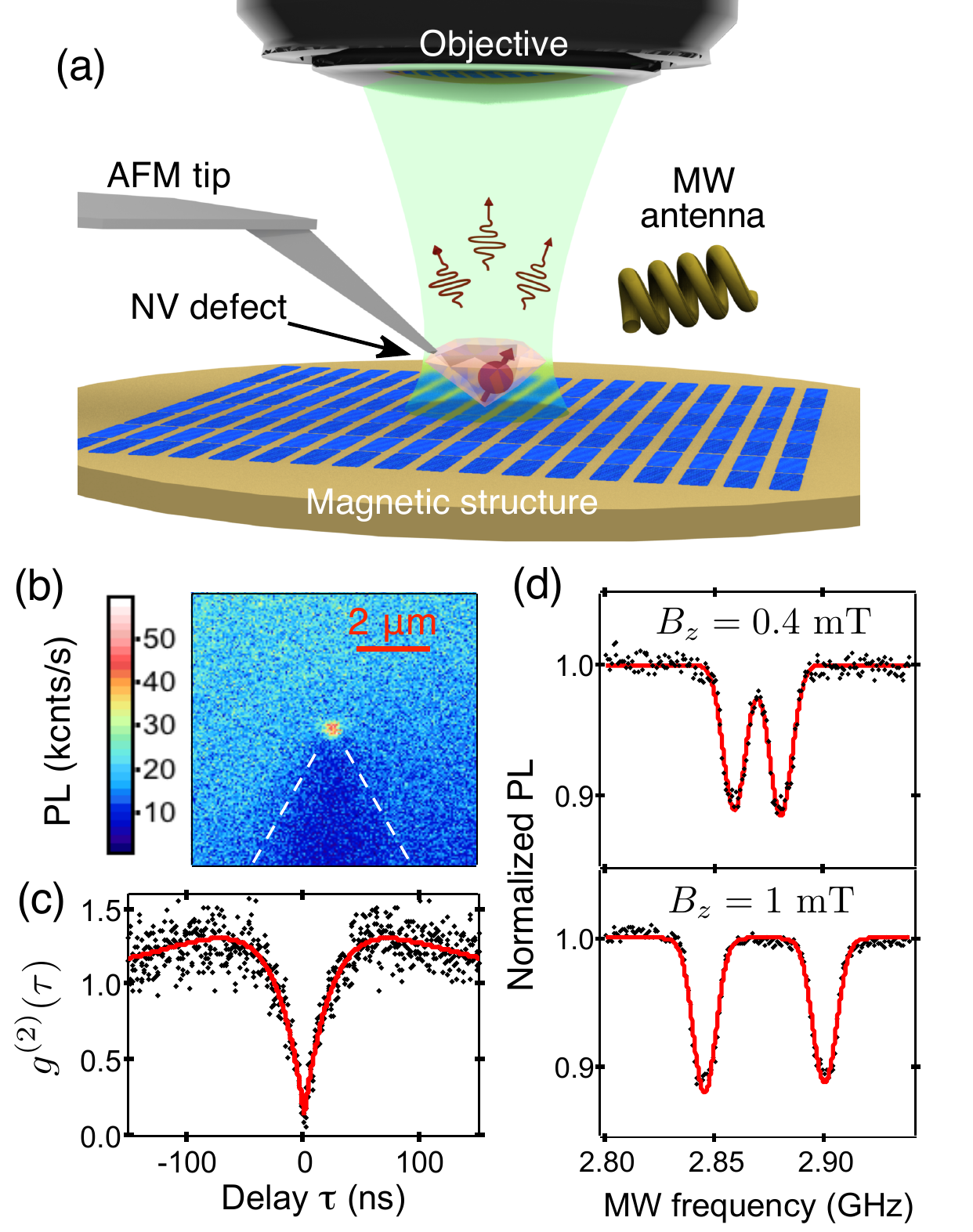}
\caption{(a)-Simplified scheme of the scanning probe magnetometer. A 20~nm diamond nanocrystal hosting a single NV defect is grafted at the end of an AFM tip. A confocal microscope placed on top of the tip allows us both to excite and collect the NV defect spin-dependent PL. A microwave field (MW) is generated by an antenna approached in the vicinity of the NV defect. (b)-PL raster scan of the AFM tip showing a bright PL spot at its apex. White dashed lines indicate the shadow of the tip. (c)- Second-order autocorrelation function $g^{(2)}(\tau)$ recorded from the bright emission spot using a Hanbury Brown and Twiss interferometer. A strong anticorrelation effect, $g^{(2)}(0) \approx 0.1$, is observed at zero delay. More than $1\times 10^{5}$ photons per second can be detected from the single NV defect when excited at saturation, with a signal-to-background ratio greater than $10$. (d)-Optically detected ESR spectra recorded for different magnetic field magnitudes. For these experiments, the ESR contrast is $\mathcal{C}\approx 12\%$, the linewidth $\Delta_{\rm ESR}\approx 9$~MHz, and the average rate of detected photons $\mathcal{R}\approx 6\times 10^{4}$ cnt/s, with an optical pumping power $\mathcal{P}_{\rm opt}=300 \ \mu$W. The zero-field splitting parameters are $D=2.87$ ~GHz and $E=5$~MHz.
}
\label{fig:setup}
\end{figure}
\indent As depicted in Fig.~\ref{fig:setup}(a), the scanning probe magnetometer combines an optical confocal microscope placed on top of a customized tuning-fork-based AFM (Attocube Systems), all operating under ambient conditions. The atomic-sized magnetic sensor was first engineered by grafting a diamond nanocrystal hosting a single NV defect at the end of the AFM tip~\cite{Sup,Cuche2009,Rondin2010}. After functionalization, a PL raster scan of the AFM tip shows an isolated spot at its apex (Fig.~\ref{fig:setup}(b)), which corresponds to the collected light emitted by a single NV defect, as verified using photon correlation measurements (Fig.~\ref{fig:setup}(c)).\\
\indent The magnetic response of the probe was first characterized by sweeping the frequency of a driving microwave (MW) field through the spin resonance while recording the NV defect PL. Owing to spin dependent PL, ESR appears as a drop of the PL signal (Fig.~\ref{fig:setup}(d))~\cite{Gruber_Science1997}. The spin Hamiltonian of the NV defect reads as~\cite{Balasubramanian2008}  
\begin{equation}
\mathcal{H}=h DS_{z}^{2}+h E(S_{x}^{2}-S_{y}^{2})+\hbar\gamma_{e}\mathbf{B}\cdot\mathbf{S} \ ,
\label{eq:h}
\end{equation}
where $h$ is the Planck constant, $D$ and $E$ are the zero-field splitting parameters, $z$ is the NV defect quantization axis, $\gamma_{e}$ the electron gyromagnetic ratio, and $\mathbf{B}$ the local magnetic field applied to the NV defect electron spin $S=1$. For magnetic fields smaller than roughly 10 mT, \textit{i.e.} such that $\hbar\gamma_e B \ll hD $, the quantization axis is fixed by the NV defect axis itself and the ESR frequencies are given by $\nu_{R}=D\pm\sqrt{(\gamma_{e}B_{z}/2\pi)^2+E^{2}}$, where $B_{z}$ is the magnetic field amplitude along the NV axis. The spin orientation of the magnetic sensor was determined by recording ESR frequencies as a function of the orientation and the magnitude of a calibrated magnetic field~\cite{Sup}.\\
\indent The shot-noise-limited sensitivity to d.c.~magnetic field $\eta_{B}$ is linked to the minimum detectable magnetic field along the NV axis $B_{z}^{\rm min}$ during an acquisition time $\Delta t$ through the relation~\cite{Dreau2011} 
\begin{equation}
\eta_B(\mathrm {T}/\sqrt{\mathrm{Hz}})=B_{z}^{\rm min}\sqrt{\Delta t}\approx \frac{1}{\gamma_{e}} \times\frac{2\pi\Delta_{\rm ESR}}{\mathcal C\sqrt{\mathcal R}} \ ,
\label{eq:sensitivity}
\end{equation}
where $\mathcal R$ is the rate of detected photons, $\mathcal C$ the ESR contrast and $\Delta_{\rm ESR}$ the associated linewidth (FWHM). From the optically detected ESR spectra shown in Fig.~\ref{fig:setup}(d), we infer $\eta_B\approx 10 \ \mu$T/$\sqrt{\rm Hz}$. We note that this sensitivity could be significantly enhanced by using pulsed-ESR techniques~\cite{Maze2008,Dreau2011} or by engineering a single NV defect in an ultra-pure diamond nanostructure~\cite{MaletinskyArxiv_2011}. \\
\indent The performances of the NV scanning probe magnetometer were characterized by mapping the magnetic field distribution created by a commercial magnetic hard disk. The magnetic bit geometry of this simple test sample is depicted in Fig.~\ref{fig:bmap}(a). As a first experiment, the magnetic sensor was approached at roughly $250$~nm from the sample surface. At such a distance, the random orientation of the magnetic bits provides non-trivial magnetic field patterns with a typical magnetic field magnitude in the mT range~\cite{Sup}. The simplest way to map such a magnetic field distribution is borrowed from magnetic resonance imaging~\cite{Balasubramanian2008}. A MW field was applied with a fixed frequency corresponding to the zero-field ESR frequency. Iso-magnetic field images were then recorded by monitoring the NV defect PL while scanning the hard disk~\cite{Balasubramanian2008}. Dark areas corresponding to zero magnetic field lines can be observed (Fig.~\ref{fig:bmap}(b)). However, the image exhibits a varying contrast related to inhomogeneous background luminescence from the sample. This limitation can be overcome by measuring the difference of PL for two fixed MW frequencies applied consecutively at each point of the scan (Fig.~\ref{fig:bmap}(c)). In this dual iso-magnetic field image, the contrast is significantly improved since background luminescence from the sample is suppressed.\\
 \begin{figure*}[t]
  \includegraphics[width = 0.95\textwidth]{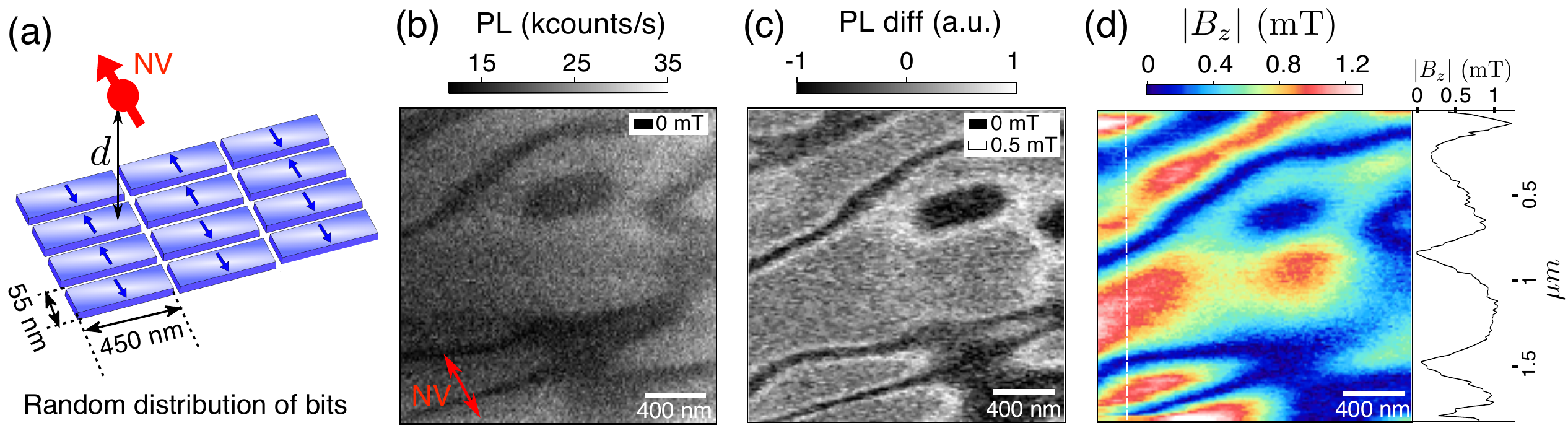}
\caption{Measurements of the magnetic field distribution created by a commercial magnetic hard disk for a probe-to-sample distance $d=250$~nm. All images correspond to $150\times138$ pixels, with a $13$-nm pixel size. (a)-Geometry of the magnetic bits. (b)-Single iso-magnetic field image. Dark areas correspond to a null $|B_{z}|$ component and the red arrow indicates the NV defect axis, along which the magnetic field projection is measured. (c)-Dual iso-magnetic field image recorded by measuring the PL difference for two fixed MW frequencies applied consecutively. Dark (resp. bright) areas correspond to $|B_{z}|=0$ (resp.~$|B_{z}|=0.5$~mT). (d)-Complete magnetic field distribution of the same area recorded by using the lock-in method. The panel on the right shows a line-cut taken along the white dashed line. For images (b), (c) and (d), the acquisition time per pixel is $5$~ms, $60$~ms and $110$~ms respectively. }
\label{fig:bmap}
\end{figure*}
\indent Even with two iso-magnetic field lines recorded at once, the information remains incomplete. To determine the full magnetic field distribution, an ESR spectrum could be measured for each pixel of the scan. However, since a few seconds are required to record such a spectrum with a reasonable signal-to-noise ratio, this method would be extremely slow. To circumvent this limitation, a lock-in method was developed in order to track the shift of the ESR frequency while scanning the magnetic sample~\cite{Schoenfeld2011}. This technique allows us to measure the strength of the magnetic field along the NV axis over the full scanning area with a typical sampling time of $110$~ms~\cite{Sup}. As shown in Fig.~\ref{fig:bmap}(d), the complete magnetic field map is in clear agreement with iso-magnetic field experiments. In addition, we note that the recorded pattern can be qualitatively reproduced by numerical simulations, both in shape, size, and magnetic field amplitude~\cite{Sup}. This confirms the reliability of NV-based scanning probe magnetometry for imaging non trivial magnetic field distributions at the nanoscale within a reasonable data acquisition time.\\
\indent From the pixel-to-pixel noise in Fig.~\ref{fig:bmap}(d), an uncertainty $\delta B\approx27$ $\mu$T is deduced, which corresponds to a sensitivity $\eta_B^{\rm exp}\approx9$ $\mu$T/$\sqrt{\mathrm{Hz}}$, given the $110$~ms sampling time~\cite{Sup}. This value is in good agreement with the shot-noise-limited value calculated using Eq.~(\ref{eq:sensitivity}). Regarding spatial resolution, the magnetic field is experienced by the NV defect electron spin wavefunction, resulting in a subnanometric probe volume. Experimentally the spatial resolution is rather limited by the pixel size (13 nm) and the positioning accuracy. The $B$-map depicted in Fig.~\ref{fig:bmap}(d) is thus assumed to closely follow the actual field distribution -- within the noise $\delta B$ -- with an overall spatial resolution of a few tens of nm. The line-cut in Fig.~\ref{fig:bmap}(d) further illustrates the ability to follow abrupt variations, up to 0.4 mT ($\gg \delta B$) between two adjacent pixels, corresponding to a field gradient of $3\times 10^4$ T/m. We emphasize the fact that no other instrument is able to quantitatively map magnetic field distributions with such a resolution.\\
\indent Yet, many material science studies are interested in the sample magnetization ${\bf  M}$ rather than the stray field ${\bf B}$. In this context, the spatial resolution is limited not only by the probe volume but also by its distance to the sample. Here for instance, simulations indicate that the magnetic bits could be resolved by using a probe-to-sample distance smaller than 40 nm (Fig.~\ref{fig:bits}(a))~\cite{Sup}. However, the magnetic field amplitude then reaches several tens of mT, including a component $B_\perp$ orthogonal to the NV defect axis~\cite{Sup}. In such a situation, the quantization axis becomes determined by the local magnetic field rather than the NV defect axis. The eigenstates of the spin Hamiltonian are therefore given by superpositions of the $m_s=0$ and $m_s=\pm1$ spin sublevels, both in the ground and excited states~\cite{Epstein2005}. As a result, optically-induced spin polarization and spin-dependent PL of the NV defect become inefficient, and the contrast of optically detected ESR vanishes. Magnetic field sensing using the previous techniques is thus not accessible in this high off-axis magnetic field regime ($>10$ mT). However, it was also shown that the NV defect PL drops by roughly $30\%$ when the magnitude of $B_\perp$ increases from $0$ to approximately $30$~mT~\cite{Epstein2005,Lai2009}. This property can be exploited to perform a direct, microwave-free, magnetic field imaging using the single NV defect sensor.\\
   \begin{figure}[b]
  \includegraphics[width = 8.8cm]{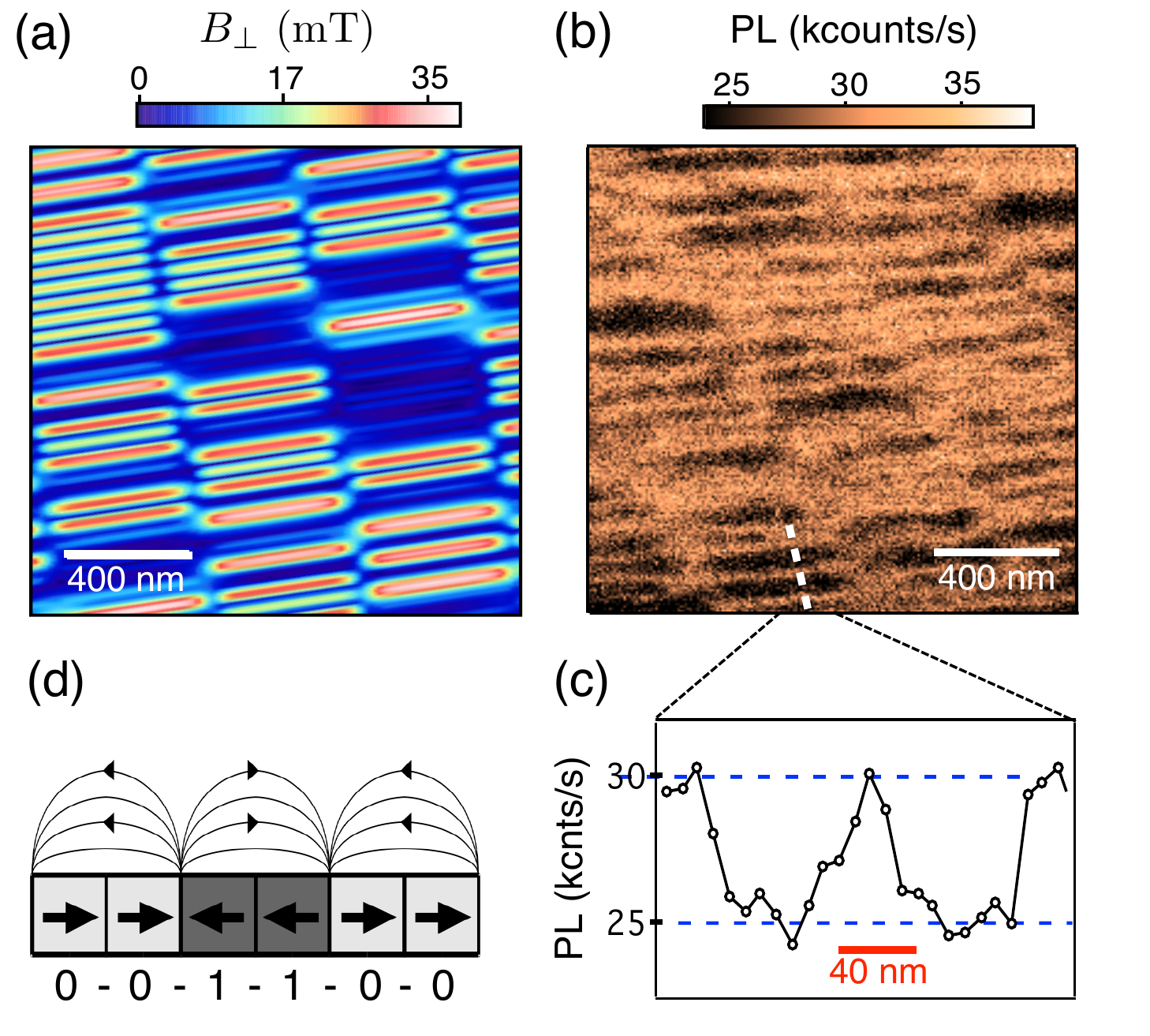}
\caption{(a)-Simulation of the off-axis magnetic field component $B_{\perp}$ for a random distribution of bit magnetizations and a probe-to-sample distance $d=30$ nm. (b)-PL image recorded with the NV scanning probe magnetometer operating in tapping mode, without applying any microwave field. The image corresponds to $200\times 200$ pixels, with a $8$-nm pixel size and a $20$-ms acquisition time per pixel. (c)-Line-cut corresponding to the white dashed line in (b). A possible configuration of the bit magnetization is schematically depicted in (d).}
\label{fig:bits}
\end{figure}
\indent  Figure~\ref{fig:bits}(b) shows a PL image recorded while scanning the magnetic sample without applying any microwave field. In this experiment, an AFM feedback loop is applied to maintain the mean probe-to-sample distance to a constant value. Comparison with simulations (Fig.~\ref{fig:bits}(a)) suggests that this distance is around 30 nm, limited by the diamond size and its position on the tip as well as by the tip oscillation. The PL image exhibits dark areas with a contrast greater than 20\% which reveal the tracks and the bits of the magnetic hard disk. We note that the contrast depends on the field strength, which itself depends on the number of bits that are consecutively parallel. For instance, from the known bit-to-bit spacing of 55 nm, we deduce that the line-cut in Fig.~\ref{fig:bits}(b) corresponds to a sequence $001100$, where $0$ and $1$ denote the two possible bit magnetizations (Fig.~\ref{fig:bits}(b)-(c)). Each bit inversion gives rise to a stray field that appears dark on the PL image. We thus demonstrate a mode of NV-based magnetic field sensing that does not require the use of microwave and is sensitive to the off-axis magnetic field magnitude. \\
\indent Summarizing, quantitative nanoscale magnetic field mapping with a sensitivity below $10$ $\mu$T$/\sqrt{\mathrm{Hz}}$ was demonstrated using NV-based scanning probe magnetometry. Moreover, we reported a microwave-free magnetic imaging technique that extends the range of application of NV-based magnetometry to large off-axis magnetic fields. Improvement of the sensitivity using pulsed-ESR techniques~\cite{Dreau2011}, together with a better control of the NV-to-sample distance, should enable imaging of few-spins systems, which would be of great interest for fundamental studies in nanomagnetism and spintronics.  

The authors acknowledge L.~Mayer, O.~Klein, F.~Mazaleyrat, M.~Lo Bue, P.~Bertet, A.~Slablab, T.~Gacoin, F.~Treussart, S.~Huant, O.~ Mollet and O.~Arcizet for fruitful discussions. This work was supported by the Agence Nationale de la Recherche (ANR) through the project D{\sc iamag}, by C'Nano \^Ile-de-France and by RTRA-Triangle de la Physique (contract 2008-057T).

\vspace{1cm}
\begin{widetext}
\section{Supplementary methods}
\subsection{Experimental setup}

The experimental setup combines an atomic force microscope (AFM) and a confocal optical microscope (AFM/CFM, Attocube System) as shown in Fig.~\ref{Fig_setup}). Two sets of piezoelectric actuators and scanners are used for positioning the sample and the AFM tip. The AFM cantilever is an Akiyama-Probe (Nanosensors) corresponding to a quartz tuning fork equipped with a micromachined silicon cantilever. The cantilever ends with a sharp tip whose nominal curvature radius is $\approx15$ nm. A nanodiamond hosting a single nitrogen-vacancy (NV) defect is placed at the end of the tip as explained in the next section.\\
\indent A continuous laser source operating at $532$~nm wavelength (Spectra-Physics Excelsior) is tightly focused at the end of the AFM tip through a $\times100$ microscope objective with a $0.9$-numerical aperture and a $1$-mm working distance (Olympus, MPLFLN100X). The photoluminescence (PL) emitted by a single NV defect placed at the apex of the tip is collected by the same objective and spectrally filtered from the remaining pump light using a dichroic beam splitter and a band-pass filter (Semrock, 697/75 BP) . Following standard confocal detection scheme, the collected light is then focused onto a $50$-$\mu$m diameter pinhole and directed to silicon avalanche photodiodes (Perkin-Elmer, SPCM-AQR-14) operating in the single-photon counting regime. Independently of the AFM scanners, precise positioning of the NV defect at the focus of the confocal microscope is achieved by scanning the laser beam with a fast steering mirror (Newport, FSM-300-02) combined with a pair of telecentric lenses. The focal length is adjusted by using the piezoelectric positioner of the AFM tip. \\
\begin{figure}[b]
\begin{center}
\includegraphics[width=9.5cm]{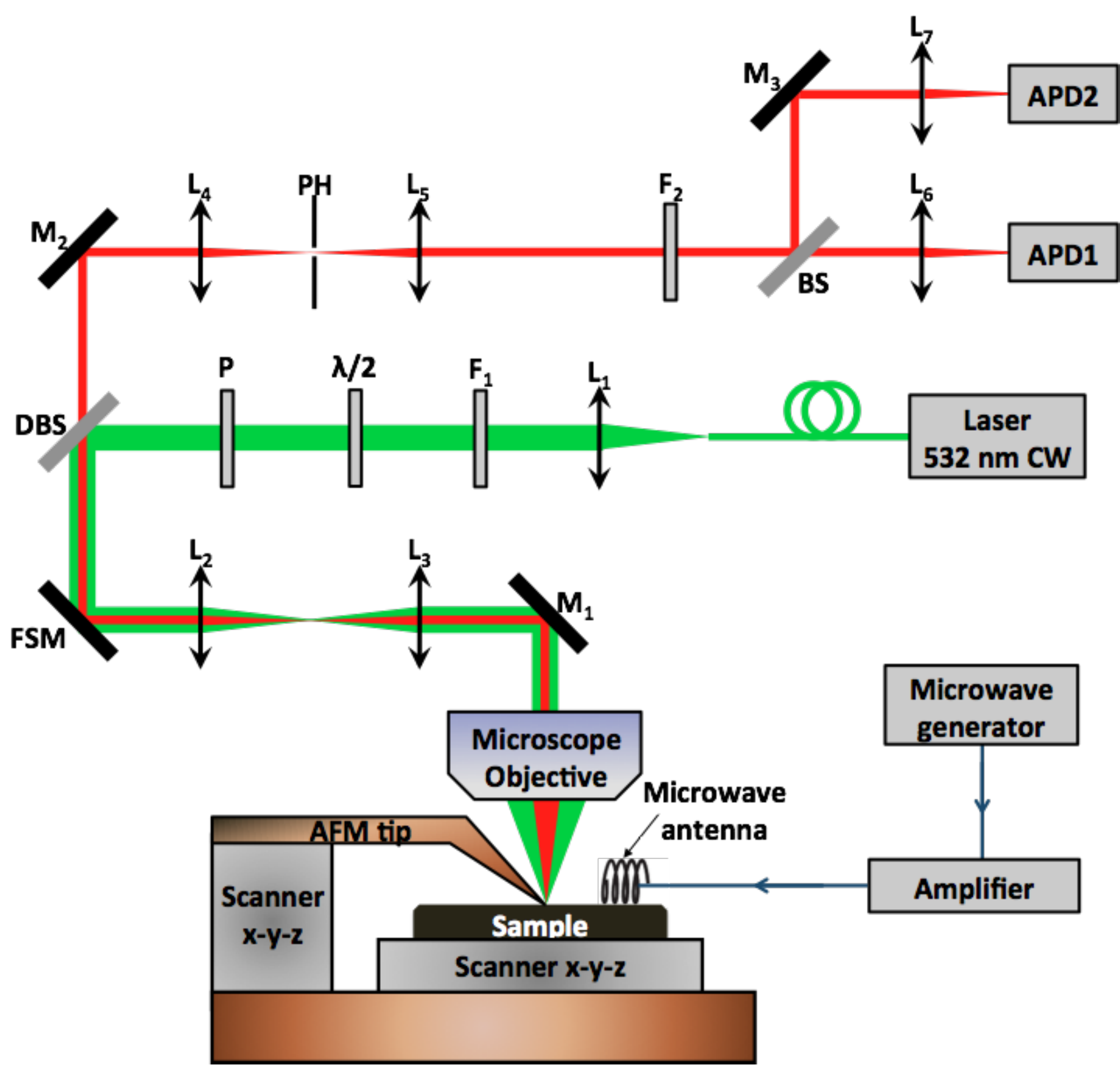}
\caption{Schematic of the experimental setup, showing both the confocal optical microscope (upper part) and the tuning-fork based AFM microscope (lower part). L stands for lens, F for filter, $\lambda/2$ for half-wave plate, P for polarizer, BS for beamsplitter, DBS for dichroic beamsplitter, M for mirror, FSM for fast steering mirror, PH for pinhole and APD for silicon avalanche photodiode. F$_1$ (resp. F$_2$) is a $10$~nm (resp. 75nm) width interference filter centered at $532$~nm (resp. 697~nm). }
\label{Fig_setup}
\end{center}
\end{figure}
\indent The unicity of the emitter was checked by using a Hanbury Brown and Twiss (HBT) interferometer consisting of two avalanche photodiodes placed on the output ports of a $50/50$ beamsplitter. The HBT detection system is associated with a fast multichannel analyser (Picoquant, TimeHarp 300) for recording the histogram of time delays between two consecutive single-photon detections. After normalization to a Poissonnian statistics, the recorded histogram is equivalent to a measurement of the second-order autocorrelation function $g^{(2)}(\tau)$~\cite{Brouri}. The observation of an anticorrelation effect at zero time delay, $g^{(2)}(0)< 0.5$, is the signature that a single NV defect is addressed (see Fig.~1(c) of the main text).\\
\indent For the electron spin resonance (ESR) measurements, a microwave synthesizer (Rohde \& Schwarz, SMR20) is followed by a power amplifier (Mini-Circuits, ZHL-42) and connected to a $20$-$\mu$m-diameter copper wire directly spanned on the sample surface.\\
\indent Scanning probe magnetometry is performed by monitoring the spin-dependent NV defect PL while scanning the magnetic sample. The tip-to-sample distance is adjusted using the piezoelectric scanner of the sample. For tapping mode operation, usual tuning-fork based AFM operation is used with a feedback loop adjusting the tip-to-sample distance.

\subsection{Grafting a single NV defect at the end of the AFM tip}
\label{graft}
We started from commercially available diamond nanocrystals (SYP 0.05, Van Moppes SA, Geneva). Such nanocrystals are produced by milling  type-Ib high-pressure high-temperature diamond crystals with a high nitrogen content ($[N]\approx 200$~ppm). The formation of NV defects was carried out using high energy ($13.6$~MeV) electron irradiation followed by annealing at $800^{\circ}$C under vacuum during two hours. The electron irradiation creates vacancies in the diamond matrix while the annealing procedure activates the migration of vacancies to intrinsic nitrogen impurities, leading to NV defect bonding.  After annealing, the irradiated nanocrystals were oxidized in air at $550^{\circ}$C during two hours. This procedure reduces the size of the nanodiamonds~\cite{Gaebel} and leads to an efficient charge state conversion of the created NV defects into the negatively-charged state~\cite{Rondin2010}. After sonication and washing with distilled water, aqueous colloidal solutions of dispersed nanodiamonds, with a mean size of 20 nm, were obtained and finally deposited by spin-coating onto a silica coverslip.\\
\indent A nanodiamond (ND) hosting a single NV defect was grafted at the apex of the AFM tip following the method introduced by Cuche {\it et al.}~\cite{Cuche2009,Cuche2010}. The tip was first dipped in a solution of poly-L-lysine (EMS, molecular weight 30000-70000 u) for a few minutes. The tip as well as the sample of dispersed nanodiamonds were then mounted in the experimental set-up, and the NDs were optically characterized. Once a ND hosting a single negatively-charged NV defect was found, the sample was scanned with the AFM tip in tapping mode in a small area around the preselected ND. Electrostatic interactions between the polymer and the ND allow reproducible and robust grafting of the nanodiamond at the apex of the AFM tip. 

\subsection{Real-time tracking of the ESR frequency}

\begin{figure}[t]
\begin{center}
    \includegraphics[width=7.5cm]{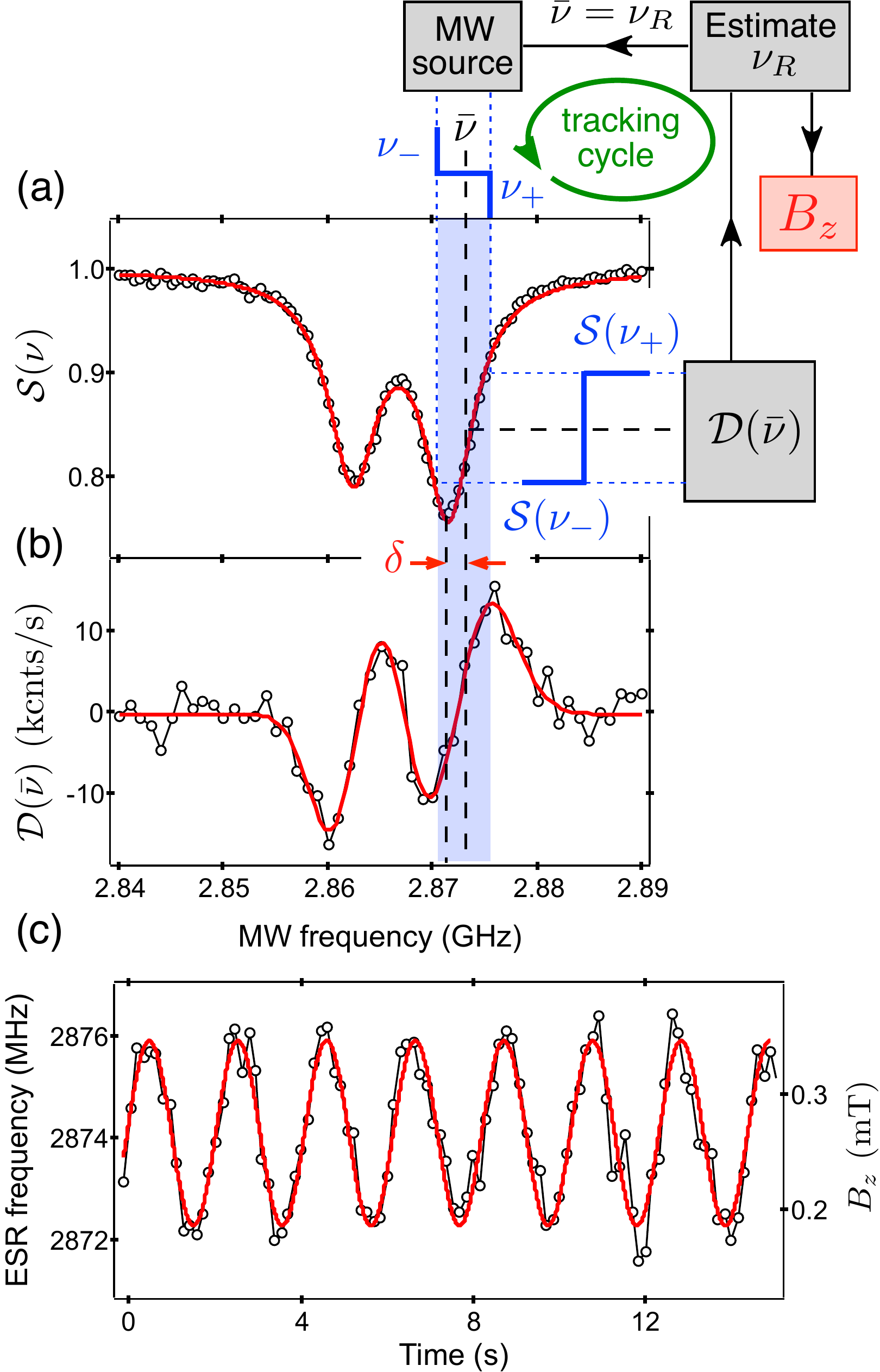}
\end{center}
\caption{Real-time tracking of the ESR frequency. (a)-Optically detected ESR spectrum $\mathcal{S}(\nu)$ and schematic of the field-frequency lock-in. (b)-Error signal $\mathcal{D}(\bar{\nu})$ used for the feedback loop. Provided that $\delta<\Delta_{\rm ESR}$, $\mathcal{D}(\bar{\nu})=\mathcal{A}\times\delta$ at first order, where $\mathcal{A}$ is experimentally determined and corresponds to the first derivative of $\mathcal{D}(\bar{\nu})$ evaluated at the resonance frequency $\nu_{R}$. For each cycle of the tracking algorithm, $\mathcal{D}(\bar{\nu})$ around the former ESR frequency is measured and the local magnetic field $B_{z}$ is deduced. (c)-Real-time measurement of the magnetic field created by a $0.5$ Hz AC current in a coil with a sampling rate of $150$ ms. Parameters of the feedback loop are $\Delta \nu=5$ MHz and $\mathcal{A}=5\times 10^3$~ cnts.s$^{-1}$.MHz$^{-1}$.}
\label{fig:lockin}
\end{figure}
The lock-in method allowing to track the shift of the ESR frequency $\nu_{R}$ while scanning the magnetic sample is described in Fig.~\ref{fig:lockin}.
Two slightly different microwave frequencies $\nu_{+}$ and $\nu_{-}$, defined by $\nu_{\pm}= \bar{\nu}\pm \Delta \nu/2$, are applied consecutively and the PL difference $\mathcal{D}(\bar{\nu})=\mathcal{S}(\nu_{+})-\mathcal{S}(\nu_{-})$ is computed (Fig.~\ref{fig:lockin}(a) and (b)). This signal can be used as an error signal for tracking the ESR frequency~\cite{Schoenfeld2011} . Indeed, if $\bar{\nu}=\nu_{R}+\delta$ with $\delta$ smaller than the ESR linewidth (FWHM), the error signal is at first order proportional to the frequency detuning $\delta$, {\it i.e.} to the magnetic field variation (Fig.~\ref{fig:lockin}(b)). Monitoring $D(\bar \nu)$ thus enables quantitative magnetic field mapping by continuously tracking the ESR frequency and measuring the frequency detuning from point to point, with a typical sampling time of $100$~ms. Reliability of this method is demonstrated by monitoring in real-time the magnetic field created by an AC current passing through a coil with a frequency up to $1$ Hz, as shown in Fig.~\ref{fig:lockin}(c).

\subsection{Estimation of the magnetic field sensitivity}

Since the magnetic stray field of the investigated hard disk is a priori not known, the measured field map (Fig.~2(c) in the main text) cannot be directly compared to theory. However, the field is assumed to vary slowly with respect to the spatial sampling, {\it i.e.} the pixel size, so that high-frequency components are attributed to measurement noise. Therefore, we introduce a pixel-to-pixel noise to estimate the experimental error $\delta B$. If $\lbrace B_{i,j}\rbrace$ is the matrix corresponding to the magnetic field map, with $1\leq i\leq p$ and $1\leq j\leq q$, then $\delta B$ can be defined as the standard deviation
\begin{equation}
\delta B^2=\frac{\sum_{i=1}^{p-1}\sum_{j=1}^{q}\left(\frac{B_{i+1,j}-B_{i,j}}{2}\right)^2
+\sum_{i=1}^{p}\sum_{j=1}^{q-1}\left(\frac{B_{i,j+1}-B_{i,j}}{2}\right)^2}{(p-1)\times q + p\times (q-1)}
\end{equation}
which takes into account the pixel-to-pixel noise along both axis. For Fig.~2(c) of the main text, for which $p=138$ and $q=150$, we find $\delta B=27$ $\mu$T. Given the acquisition time per pixel $\Delta t= 110$ ms, an experimental field sensitivity $\eta_B^{exp}=\delta B \sqrt{\Delta t}=9$ $\mu$T$/\sqrt{\mathrm{Hz}}$ is obtained.

\subsection{Determination of the NV defect axis}

The NV defect axis, characterized by the unit vector $\hat{z}$, can be determined by measuring the ESR spectrum while applying a static magnetic field along several well-chosen directions. Denoting $\theta$ and $\phi$ the polar and azimuthal angles of the applied magnetic field in the reference frame $xyz$ of the NV defect, the diagonalization of the electron spin Hamiltonian (equation~(1) of the main text) yields to the relation~\cite{Balasubramanian2008} 
\begin{eqnarray} \label{eqn_theta}
\Delta = \frac{7D^3+2(\nu_1+\nu_2)\left(2\left(\nu_1^2+\nu_2^2\right)-5\nu_1\nu_2-9E^2\right)-3D\left(\nu_1^2+\nu_2^2-\nu_1\nu_2+9E^2\right)}{9\left(\nu_1^2+\nu_2^2-\nu_1\nu_2-D^2-3E^2\right)} \ ,
\end{eqnarray}
where $\nu_1$ and $\nu_2$ are the Zeeman-shifted ESR frequencies, $\Delta=D \cos 2\theta+2E\cos 2\phi\sin^2\theta$ while $D$ and $E$ are the zero-field splitting parameters of the NV ground-state electron spin. Since $E \ll D$, one can use the approximation $\Delta\approx D\cos2\theta$. \\
\indent For a magnetic field applied along a given direction $\hat{K}$, Eq. (\ref{eqn_theta}) gives the angle $|\theta_K|$ between $\hat{z}$ and $\hat{K}$. Repeating the measurement for two other directions of the magnetic field enables to completely determine $\hat{z}$. Since the magnetic images presented in the main manuscript are $XY$ maps of the magnetic field component $|B_z|$, it is important to know $\hat{z}$ with respect to the reference frame $XYZ$ of the laboratory. The orientation of the NV defect used in Fig. 2 and 3 of the main text was determined following the latter method. Fig. \ref{Fig_orientation}(b) shows the evolution of the measured resonance frequencies $\nu_1$ and $\nu_2$ as a function of a static magnetic field produced by two coils in an Helmholtz configuration. This field was first aligned along the $X$-axis (filled circles) and then along the $Y$-axis (open circles). From these measurements and Eq. (\ref{eqn_theta}), the angles $|\theta_X|=37\pm1^\circ$ and $|\theta_Y|=53\pm1^\circ$ were obtained. As $|\theta_X|+|\theta_Y|=90^\circ$, one deduces that $\hat{z}$ lies in the $XY$ plane. The actual orientation of the NV axis in the reference frame $XYZ$ is illustrated in Fig.~\ref{Fig_orientation}(a). The other orientation compatible with the set ($|\theta_X|$,$|\theta_Y|$) was ruled out by a measurement along a third direction.
\begin{figure}[h!] 
\begin{center}
\includegraphics[width=10cm]{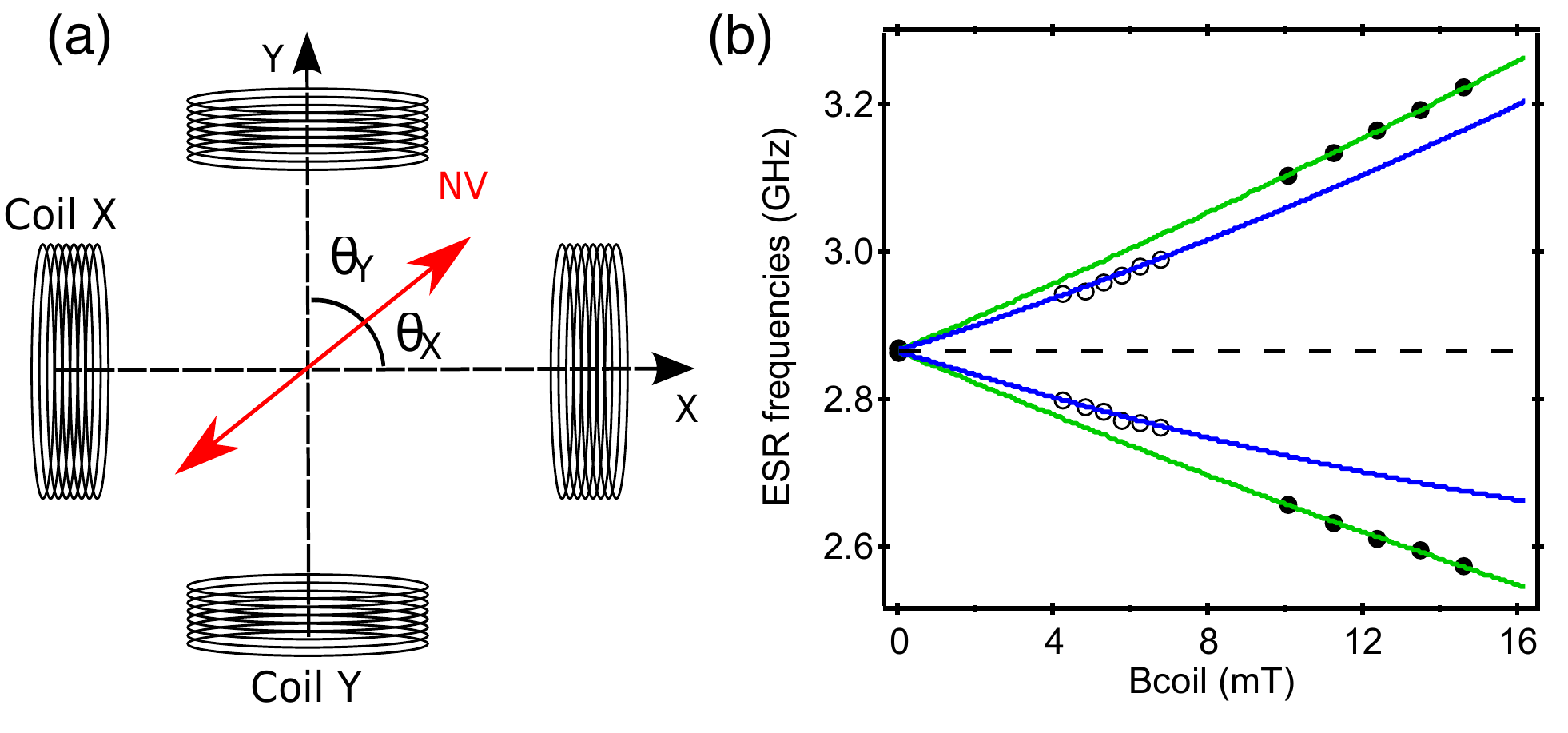}
\caption{Measurement of the NV defect orientation using Helmholtz coils. (a)-Measuring the ESR spectrum while applying a static magnetic field along $X$ (resp. $Y$) gives access to the angle $|\theta_X|$ (resp. $|\theta_Y|$) between the NV axis (red arrow) and the $X$-axis (resp. $Y$-axis). (b)-ESR frequencies $\nu_{1}$ and $\nu_{2}$ as a function of the magnetic field magnitude. The filled (resp. open) black circles are the experimental data for a static magnetic field applied along $X$ (resp. $Y$). The angles $|\theta_X|$ and $|\theta_Y|$ were estimated using Eq. (\ref{eqn_theta}). The solid lines are the theoretical curves using the full diagonalization of the spin Hamiltonian and the calculated values of $|\theta_{X}|$ and $|\theta_{X}|$, considering a field oriented along $X$ (green lines) and $Y$ (blue lines), respectively.}
\label{Fig_orientation}
\end{center}
\end{figure}

\section{Modelling of the magnetic field distribution above the hard disk}

The magnetic sample used for the demonstration of magnetic field mapping is a piece of a commercial magnetic hard disk (Maxtor, 20~Gb). The geometry of the magnetic bits is depicted in Fig.~\ref{Fig_S4}. AFM topography of the sample allows to infer the geometrical orientation of the bits since grooves imprinted by the read/write head indicate the direction of the tracks (Fig.~\ref{Fig_S4}(a)). This information is used to determine the NV orientation with respect to the bits orientation.\\
\indent The hard disk was modeled as an array of uniformly magnetized parallelepipeds (the `bits'). The dimensions of each `bit' are $2a=40$ nm, $2b=400$ nm and $2c=10$ nm, along $X$, $Y$ and $Z$, respectively, and their magnetization is ${\bf M}=\pm M_s \hat{X}$ depending whether the bit is at `0' or `1' (Fig.~\ref{Fig_S4}(b)-(c)). $M_s$ is the average magnetization of the ferromagnetic material. Typical value for the studied hard disk is  $\mu_0 M_s=0.5$~T where $\mu_0$ is the permeability of free space. The pitch of the array is 55 nm along $X$ and 450 nm along $Y$. \\
\begin{figure}[b] 
\centerline{\includegraphics[width=16cm]{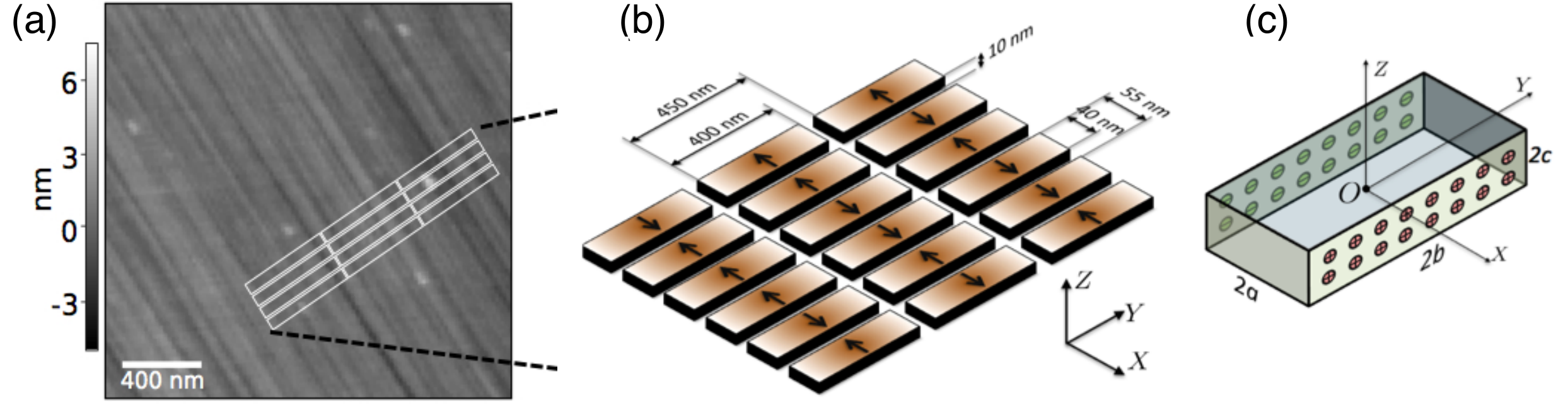}}
\caption{Modelling the magnetic hard disk. (a)-AFM Topography of a $2\times 2 \ \mu$m area of the hard disk. (b)-Model used for the $20$~Gb hard disk. Each bit is magnetized along $\pm \hat{X}$ and has dimensions 400 x 40 x 10 nm. The bits are spaced by 450 nm along $Y$ (track width) and 55 nm along $X$. (c)- A parallelepiped uniformely magnetized along $X$ is equivalent to a magnetic ``capacitor'' with two charged surfaces located at $X=\pm a$.}
\label{Fig_S4}
\end{figure}

\indent The magnetic field distribution produced by a single bit (Fig.~\ref{Fig_S4}(c)) can be calculated using magnetostatics theory. Outside the parallelepiped, the three components of the magnetic field read as 
\begin{figure}[b] 
\centerline{\includegraphics[width=15.5cm]{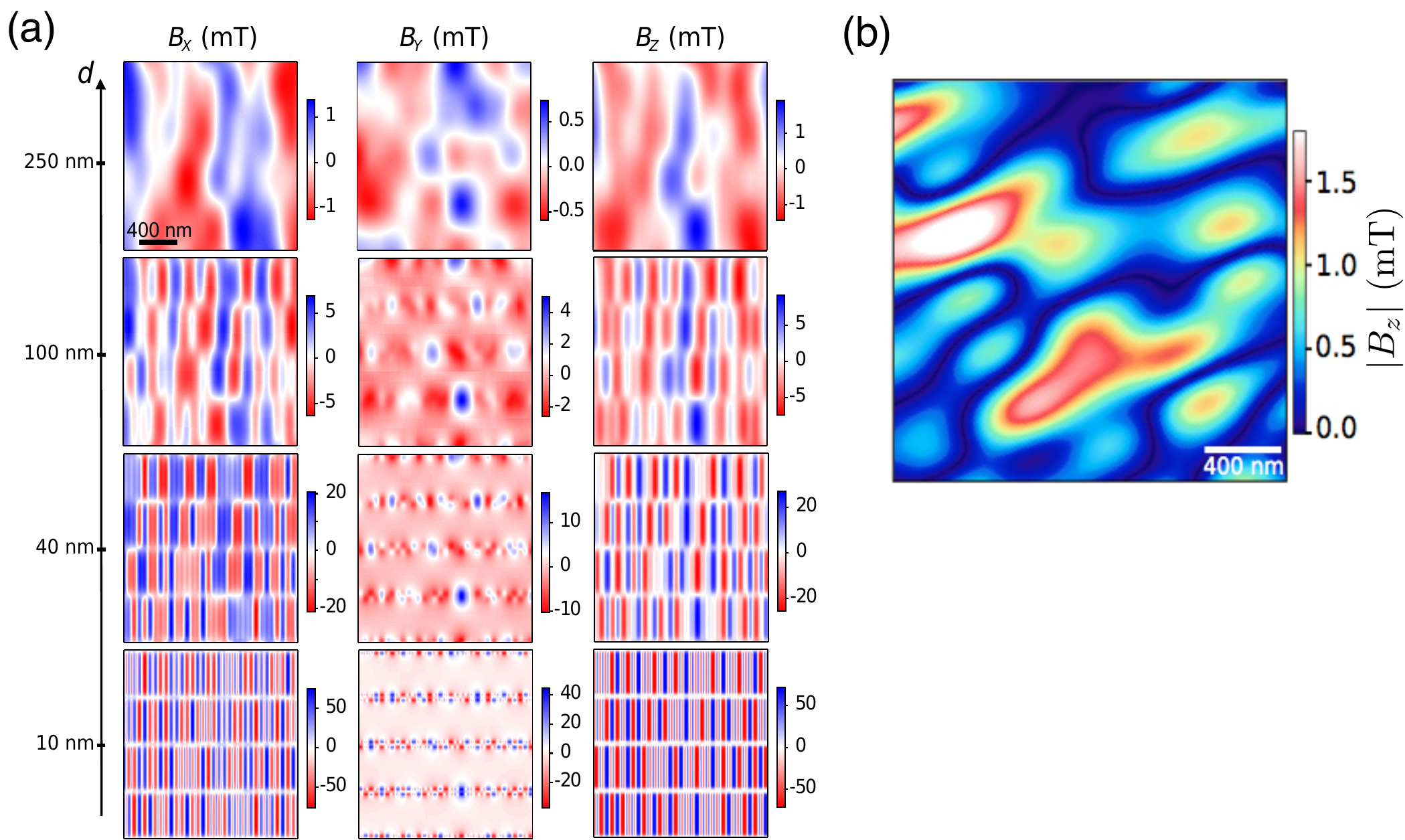}}
\caption{(a)-Calculated magnetic field distributions above a randomly written area of the hard disk. The three components $B_X$, $B_Y$ and $B_Z$ (columns 1, 2 and 3, respectively) are calculated at various distances $d$ from the surface, from $10$ nm to $250$ nm (from bottom to top). The plots are 1.8 x 1.8 $\mu$m in size, but the array used for the calculation is 9 times larger (5.4 x 5.4 $\mu$m) in order to avoid edge effects. (b)-Simulation of the magnetic field created by a random distribution of bit magnetizations for $d=250$~nm, once projected along the NV defect axis used in the experiments described in the main text of the article. The magnetic bits are oriented as shown in the AFM topography (Fig. 4(a)).}
\label{Fig_sims}
\end{figure} 

\begin{eqnarray} \label{eqnC2} 
  B_X=-\frac{\mu_0M_s}{4\pi}\sum_{i=\pm1}\sum_{j=\pm1}\sum_{k=\pm1}ijk\tan^{-1}\frac{(Y+jb)(Z+kc)}{(X+ia)\sqrt{(X+ia)^2+(Y+jb)^2+(Z+kc)^2}}
\end{eqnarray}
\begin{eqnarray} \label{eqnC3} 
  B_Y=-\frac{\mu_0 M_s}{8\pi}
  \ln\left[\prod_{i=\pm1}\prod_{j=\pm1}\prod_{k=\pm1}
  \left(\frac{(Z+kc)-\sqrt{(X+ia)^2+(Y+jb)^2+(Z+kc)^2}}
       {(Z+kc)+\sqrt{(X+ia)^2+(Y+jb)^2+(Z+kc)^2}}\right)^{ijk}
  \right]    
\end{eqnarray}
\begin{eqnarray} \label{eqnC4}
  B_Z=-\frac{\mu_0 M_s}{8\pi}
  \ln\left[\prod_{i=\pm1}\prod_{j=\pm1}\prod_{k=\pm1}
  \left(\frac{(Y+jb)-\sqrt{(X+ia)^2+(Y+jb)^2+(Z+kc)^2}}
       {(Y+jb)+\sqrt{(X+ia)^2+(Y+jb)^2+(Z+kc)^2}}\right)^{ijk} 
  \right] \ .
\end{eqnarray}

\indent The magnetic fields produced by each bit of the array were then summed up in order to compute the whole field distribution above the hard disk. Since the exact distribution of magnetic bit orientations is unknown, we compute the magnetic field distributions along $X$, $Y$ and $Z$ directions for a random distribution of magnetic bits (Figure~\ref{Fig_sims}(a)). Figure~\ref{Fig_sims}(b) indicates the magnetic field distribution at $250$ nm from the hard disk, once projected along the axis of the NV defect used in the scanning-probe magnetometry experiments. The magnetic field follows rounded non trivial patterns with characteristics size and amplitude similar to the one observed experimentally (Fig. 2(c) of the main  text). We note that the magnetic field distribution at another distance or along the $Y$ axis follows a clearly different pattern.

\end{widetext}


\begin{thebibliography}{30}

\bibitem{Kirtley2010} 
J. R. Kirtley, Rep. Prog. Phys. \textbf{73}, 126501 (2010).

\bibitem{Romalis_NatPhys2007}
D. Budker and M. Romalis, Nature Phys. {\bf 3}, 227 (2007).	

\bibitem{Bogani2008}
L. Bogani and W. Wernsdorfer, Nature Mater. {\bf 7}, 179-186 (2008).

\bibitem{Heinze_Science2000}
S. Heinze,  M. Bode, A. Kubetzka, O. Pietzsch, X.  Nie, S. Blügel, R. Wiesendanger, Science {\bf 288}, 1805 (2000).

\bibitem{Rugar_Nature2004}
D. Rugar, R. Budakian, H. J. Mamin, and B. W.  Chui, Nature {\bf 430}, 329-332 (2004).

\bibitem{Yacoby_NanoLett2010}
A. Finkler, Y. Segev, Y. Myasoedov, M. L. Rappaport, L. Ne’eman, D.  Vasyukov, E. Zeldov, M. E. Huber, J. Martin and A. Yacoby§ Nano Lett. {\bf 10}, 1046-1049 (2010).

\bibitem{Chao2009}
W. Chao, J. Kim, S. Rekawa, P. Fischer, and E. H. Anderson, Opt. Exp. {\bf 17}, 17669 (2009).

\bibitem{MFM} 
Y. Martin and H. K. Wickramasinghe, Appl. Phys. Lett. {\bf 50}, 1455 (1987).

\bibitem{Garcia2001} 
J. M. Garcia, A. Thiaville, J. Miltat, K. J. Kirk, J. N. Chapman, and F. Alouges, Appl. Phys. Lett. \textbf{79}, 656 (2001).

\bibitem{Chernobrod2005}  
B. M. Chernobrod and G. P. Berman, J. Appl. Phys. \textbf{97}, 014903 (2005).

\bibitem{Degen_2008}
C. L. Degen, Appl. Phys. Lett. {\bf 92}, 243111 (2008).

\bibitem{Taylor2008} 
J. M. Taylor, P. Cappellaro, L. Childress, L. Jiang, D. Budker, P. R. Hemmer, A. Yacoby, R. Walsworth and M. D. Lukin, Nature Phys. \textbf{4}, {810-816} (2008).

\bibitem{Balasubramanian2008} 
G. Balasubramanian, I. Y. Chan, R. Kolesov, M. Al-Hmoud, J. Tisler, C. Shin, C. Kim, A. Wojcik, P. R. Hemmer, A. Krueger, T. Hanke, A. Leitenstorfer, R. Bratschitsch, F.  Jelezko and J. Wrachtrup, Nature \textbf{455}, 648-651 (2008).

\bibitem{Maze2008} 
J. R. Maze, P. L. Stanwix, J. S. Hodges, S. Hong, J. M. Taylor, P. Cappellaro, L. Jiang, M. V. Gurudev Dutt,  E. Togan, A. S. Zibrov, A. Yacoby, R. L. Walsworth and M. D. Lukin, Nature \textbf{455}, 644-647 (2008).

\bibitem{Sup}
See supplementary material at [URL will be inserted by AIP] for experimental details and numerical simulations of the magnetic field distribution above the magnetic hard disk.

\bibitem{Cuche2009} 
A. Cuche, A.  Drezet, Y. Sonnefraud, O. Faklaris, F. Treussart, J.-F. Roch, and S. Huant , Opt. Exp. {\bf 17}, 19969 (2009).

\bibitem{Rondin2010} 
L. Rondin, G. Dantelle, A. Slablab, F. Grosshans, F. Treussart, P. Bergonzo, S. Perruchas, T. Gacoin, M. Chaigneau, H.-C. Chang, V. Jacques and J.-F. Roch , Phys. Rev. B \textbf{82}, 115449 (2010).

\bibitem{Gruber_Science1997} 
A. Gr$\ddot{\rm u}$ber, A. Drabenstedt, C. Tietz, L. Fleury, J. Wrachtrup, and C. von Borczyskowski, Science \textbf{276}, 2012-2014 (1997).

\bibitem{Dreau2011}
A. Dr$\acute{\rm e}$au, M. Lesik, L. Rondin, P. Spinicelli, O. Arcizet, J.-F. Roch and V. Jacques, Phys. Rev. B, \textbf{84} 195204 (2011)


\bibitem{MaletinskyArxiv_2011}
P. Maletinsky, S. Hong, M.S. Grinolds, B. Hausmann, M.D.Lukin, R.-L. Walsworth, M. Loncar and A. Yacoby, {\it Preprint arXiv:1108.4437}.

\bibitem{Schoenfeld2011} 
R. S. Schoenfeld and W. Harneit, Phys. Rev. Lett. \textbf{106}, 030802 (2011).

\bibitem{Epstein2005} 
R. J. Epstein, F. M. Mendoza, Y. K. Kato, and D. D. Awschalom, Nature Phys. \textbf{1}, 94-98 (2005).

\bibitem{Lai2009} 
N. D. Lai, D. Zheng, F. Jelezko, F. Treussart, and J.-F. Roch, Appl. Phys. Lett. \textbf{95}, 133101 (2009). 


\end{thebibliography}

\begin{thebibliography}{30}

\bibitem{Brouri}
R. Brouri, A. Beveratos, J.-P. Poizat,  and P. Grangier, Opt. Lett. {\bf 25}, 1294 (2000). 

\bibitem{Gaebel}
T. Gaebel, C. Bradac, J. Chen, P. Hemmer, and J. R. Rabeau, preprint arXiv:1104.5075 (2011). 

\bibitem{Rondin2010} 
L. Rondin {\it et al.}, Phys. Rev. B \textbf{82}, 115449 (2010).

\bibitem{Cuche2009} 
A. Cuche {\it et al.}, Opt. Exp. {\bf 17}, 19969 (2009).

\bibitem{Cuche2010} 
A. Cuche, A. Drezet, J.-F. Roch, F. Treussart, and S. Huant, J. Nanophoton. {\bf 4}, 043506 (2010).

\bibitem{Balasubramanian2008} 
G. Balasubramanian {\it et al.}, Nature \textbf{455}, 648-651 (2008).

\bibitem{Garcia2001} 
J. M. Garcia, A. Thiaville, J. Miltat, K. J. Kirk, J. N. Chapman, and F. Alouges, Appl. Phys. Lett. \textbf{79}, 656 (2001).



\end{thebibliography}
\end{document}